\documentclass[sigconf]{acmart}

\renewcommand\footnotetextcopyrightpermission[1]{}
\usepackage{paralist}
\usepackage{enumitem}
\usepackage{multirow}
\usepackage[table]{xcolor}
\usepackage{colortbl}
\usepackage[most]{tcolorbox}
\usepackage{tabularx}
\usepackage{soul}
\definecolor{promptblue}{RGB}{88, 103, 154}
\definecolor{promptgray}{RGB}{242, 242, 242} 
\newtcolorbox{promptbox}[1]{
	colback=promptgray,
	colframe=promptblue,
	coltitle=white,
	title={#1},
	arc=4pt,
	boxrule=1pt,
	fonttitle=\normalfont,
	left=6pt, right=6pt, top=6pt, bottom=6pt 
}
\newtcolorbox{rqbox}{
    enhanced,
    breakable,        
    colback=promptgray,
    colframe=black,
    boxrule=0.5pt,
    arc=1mm,
    left=4pt,
    right=4pt,
    top=4pt,
    bottom=4pt
}
\newcommand{\originhl}[1]{\colorbox{gray!25}{\strut #1}}
\newcommand{\triphl}[1]{\colorbox{purple!18}{\strut #1}}
\newcommand{\lpraghl}[1]{\colorbox{orange!20}{\strut #1}}
\newcommand{\parahl}[1]{\colorbox{green!18}{\strut #1}}
\newcommand{\redacthl}[1]{\colorbox{blue!15}{\strut #1}}
\newcommand{\origintexthl}[1]{{\sethlcolor{gray!25}\hl{#1}}}
\newcommand{\triptexthl}[1]{{\sethlcolor{purple!18}\hl{#1}}}
\newcommand{\lpragtexthl}[1]{{\sethlcolor{orange!20}\hl{#1}}}
\newcommand{\paratexthl}[1]{{\sethlcolor{green!18}\hl{#1}}}
\newcommand{\redacttexthl}[1]{{\sethlcolor{blue!15}\hl{#1}}}
\AtBeginDocument{%
  }

\setcopyright{acmlicensed}
\copyrightyear{2018}
\acmYear{2018}
\acmDOI{XXXXXXX.XXXXXXX}
\acmConference[Conference acronym 'XX]{Make sure to enter the correct
  conference title from your rights confirmation email}{June 03--05,
  2018}{Woodstock, NY}
\acmISBN{978-1-4503-XXXX-X/2018/06}

\begin{document}

\title[Not All Entities are Created Equal]{Not All Entities are Created Equal: A Dynamic Anonymization Framework for Privacy-Preserving RAG}


\author{Xinyuan Zhu}
\authornote{Both authors contributed equally to this research.}
\orcid{0009-0003-7177-9373}
\affiliation{%
  \institution{Nankai University}
  \city{Tianjin}
  \country{China}
}
\email{zhuxinyuan@mail.nankai.edu.cn}

\author{Zekun Fei}
\authornotemark[1]
\affiliation{%
  \institution{Nankai University}
  \city{Tianjin}
  \country{China}
}
\email{feizekun@mail.nankai.edu.cn}

\author{Enye Wang}
\affiliation{%
  \institution{Nankai University}
  \city{Tianjin}
  \country{China}}
\email{wangenye@mail.nankai.edu.cn}

\author{Ruiqi He}
\affiliation{%
  \institution{Nankai University}
  \city{Tianjin}
  \country{China}
}
\email{heruiqi@mail.nankai.edu.cn}

\author{Jia Guo}
\affiliation{%
  \institution{Gordon College}
  \country{Wenham, United States}
}
\email{jia.guo@gordon.edu}

\author{Ruijie Wang}
\affiliation{%
  \institution{Beihang University}
  \city{Beijing}
  \country{China}
}
\email{ruijiew@buaa.edu.cn}

\author{Zheli Liu}
\affiliation{%
 \institution{Nankai University}
 \city{Tianjin}
 \country{China}}
\email{liuzheli@nankai.edu.cn}

\author{Qingkai Zeng}
\authornote{Corresponding author}
\affiliation{%
	\institution{Nankai University}
	\city{Tianjin}
	\country{China}}
\email{qingkai.zeng@nankai.edu.cn}
\renewcommand{\shortauthors}{Zhu et al.}

\begin{abstract}
    Retrieval-Augmented Generation (RAG) enhances the utility of Large Language Models (LLMs) by retrieving external documents. Since the knowledge databases in RAG are predominantly utilized via cloud services, private data in sensitive domains such as finance and healthcare faces the risk of personal information leakage. Thus, effectively anonymizing knowledge bases is crucial for privacy preservation. Existing studies equate the privacy risk of text to the linear superposition of the privacy risks of individual, isolated sensitive entities. The "one-size-fits-all" full processing of all sensitive entities severely degrades utility of LLM. To address this issue, we introduce a dynamic anonymization framework named \textbf{TRIP-RAG}. Based on context-aware entity quantification, this framework evaluates entities from the perspectives of marginal privacy risk, knowledge divergence, and topical relevance. It identifies highly sensitive entities while trading off utility, providing a feasible approach for variable-intensity privacy protection scenarios. Our theoretical analysis and experiments indicate that TRIP-RAG can effectively reduce context inference risks. Extensive experimental results demonstrate that, while maintaining privacy protection comparable to full anonymization, TRIP-RAG's Recall@k decreases by less than 35\% compared to the original data, and the generation quality improves by up to 56\% over existing baselines.
\end{abstract}

\maketitle

\section{Introduction}
\label{sc:1}
\begin{figure*}[htbp]
	\centering
	\includegraphics[width=0.9\linewidth]{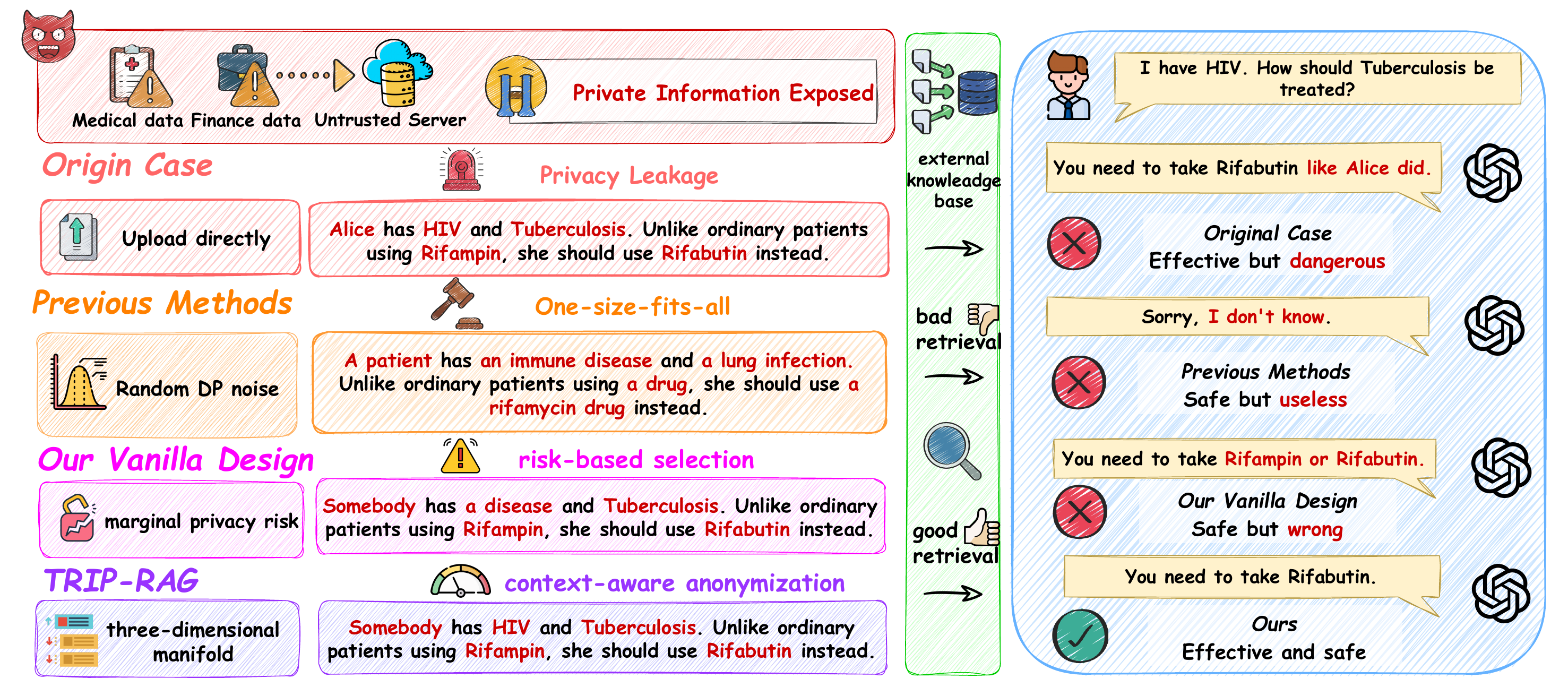} 
	\caption{Motivation of TRIP-RAG for privacy-preserving RAG. Raw sensitive data in external knowledge bases may expose private information. Uniform anonymization protects privacy but often removes utility-critical entities. TRIP-RAG selectively anonymizes entities by jointly assessing marginal privacy risk, knowledge divergence, and topical relevance, leading to better privacy-utility trade-offs.} 
	\label{fig:1}
\end{figure*}
While Large Language Models (LLMs) have demonstrated remarkable capabilities in language understanding and generation~\cite{das2024improved}, their knowledge is primarily acquired from large-scale publicly available corpora~\cite{tirumala2023d4}. However, publicly available corpora are unlikely to fully cover the specialized and non-public knowledge required in professional domains such as healthcare ~\cite{goyal2024healai}, finance~\cite{jin2026finrpt}, and legal services~\cite{doyle2025if}. As a result, LLMs may still struggle with tasks that require domain-specific evidence, up-to-date factual support, or professional expertise beyond their pre-training data. Retrieval-Augmented Generation (RAG)~\cite{lewis2020retrieval} mitigates this limitation by retrieving relevant documents from external knowledge bases and conditioning generation on the retrieved evidence. In this way, RAG enables LLMs to incorporate specialized knowledge that is absent from pretraining, thus improving their effectiveness in professional applications. However, this benefit often depends on indexing documents from private or regulated domains, where useful knowledge is tightly coupled with sensitive information. In such scenarios, knowledge bases may contain identifiable entities such as patient attributes, financial records, legal case information, personal identifiers, or other confidential details. 

Once such documents are indexed and retrieved in RAG systems, sensitive information may be exposed not only through the retrieved passages themselves, but also through generated responses conditioned on these passages. 
Privacy leakage arises from contextual associations, where the sensitivity of an entity depends on its surrounding text and its links to other entities~\cite{gianini2008game}. This context-dependent leakage risk is further amplified in practical RAG deployments. As illustrated in Figure~\ref{fig:1}, directly uploading raw private-domain documents to an untrusted cloud-based knowledge base may expose sensitive information to third-party infrastructure~\cite{chen2025towards,li2023chatdoctor}, making such deployments difficult to reconcile with privacy regulations such as GDPR~\cite{eu2016gdpr}. Even when the RAG service provider is assumed to be trustworthy, adversarial prompts may still induce the model to reveal raw or sensitive content from the underlying knowledge base~\cite{qi2024follow}.Therefore, privacy-preserving RAG requires sensitive information to be identified and protected before documents are indexed, retrieved, or used for generation.

Recent studies have made preliminary efforts to protect textual privacy in RAG knowledge bases, mainly through data synthesis and perturbation-based approaches. For example, \citet{zeng2025mitigating} employ adversarial LLM rewriting to synthesize privacy-preserving versions of knowledge-base texts, while \citet{he2025mitigating} propose a local differential privacy mechanism to perturb sensitive entities. Existing methods typically treat detected entities as independent units and apply uniform protection to them. Such a ``one-size-fits-all'' strategy is insufficient for privacy-preserving RAG for two reasons. First, it ignores the heterogeneous utility of different entities in RAG systems: some entities are central to document retrieval and response generation, whereas others have little impact on downstream task performance. Second, it fails to model the context-dependent nature of privacy leakage, where the risk of an entity should be assessed according to its contribution to the privacy exposure of the surrounding text rather than by its surface form alone, since anonymizing more privacy-sensitive entities in the surrounding text may alter the context such that the entity itself no longer contributes to privacy leakage. \textbf{Consequently, uniformly perturbing or removing all entities may provide strong privacy protection, but it can also destroy essential semantic and topical information, leading to substantial degradation in retrieval quality and generation utility.}

To address these limitations, we propose \textbf{TRIP-RAG}, a context-aware anonymization framework designed for privacy-preserving RAG over large-scale knowledge bases. 
Instead of anonymizing all detected entities indiscriminately, TRIP-RAG selectively protects entities by jointly considering their contextual privacy risks and utility contributions. Specifically, TRIP-RAG assesses each candidate entity from three complementary perspectives: \textbf{(1) Marginal privacy risk}, which measures how masking an entity changes the overall privacy exposure of the surrounding text using a language-model-based surrogate function trained on a small set of annotated samples; \textbf{(2) Knowledge divergence}, which compares the representations of the original document and its masked counterpart to quantify the semantic shift caused by anonymization; and \textbf{(3) Topical relevance}, which treats each entity as a proxy query and measures its distance to the document representation to estimate its importance for RAG retrieval. Entities whose content is more strongly aligned with the document topic or whose anonymization causes greater semantic deviation are more likely to contribute to retrieval and generation quality, and therefore should be anonymized more cautiously. After obtaining these three measurements, TRIP-RAG integrates them into a unified entity scoring function to guide anonymization decisions. Entities with high privacy risk and low utility contribution are prioritized for anonymization, while utility-critical entities are retained unless they pose substantial privacy risks. By adjusting the anonymization threshold over the ranked entities, TRIP-RAG can flexibly control the privacy-utility trade-off under different deployment requirements.

In summary, TRIP-RAG provides a context-aware entity anonymization mechanism that jointly considers privacy risk, semantic preservation, and retrieval utility. \textbf{This design enables the framework to protect truly privacy-revealing entities while avoiding unnecessary anonymization of utility-critical information.} Experimental results demonstrate that TRIP-RAG achieves privacy protection comparable to full anonymization while preserving substantially more retrieval and generation utility, demonstrating the effectiveness of context-aware anonymization for privacy-preserving RAG. 

The main contributions of this paper are summarized as follows:
\begin{itemize}
    \item We propose \textbf{TRIP-RAG}, a context-aware entity anonymization framework for secure RAG over large-scale knowledge bases. It selectively anonymizes privacy-revealing entities instead of masking all detected entities, thus reducing unnecessary utility loss.

    \item We design an entity scoring mechanism that jointly considers \textit{marginal privacy risk}, \textit{knowledge divergence}, and \textit{topical relevance}. It captures contextual privacy risk while estimating each entity's impact on semantic preservation and retrieval utility.

    \item We conduct extensive experiments across datasets and backbone models. Results show that TRIP-RAG achieves privacy protection comparable to full anonymization while preserving much more retrieval and generation utility, with Recall@k improving by 9\%-60\%, BLEU by 3\%-56\%, and ROUGE-L by 5\%-53\%.
\end{itemize} 

\section{Related Works}
\subsection{Privacy Exposure in RAG}
Retrieval-Augmented Generation (RAG) enhances LLMs by incorporating external non-parametric knowledge into generation~\cite{lewis2020retrieval}. By retrieving relevant documents from an external corpus and conditioning responses on retrieved evidence, RAG has been shown to improve factual accuracy, mitigate hallucinations, and support various knowledge-intensive tasks~\cite{siriwardhana2023improving,gao2023retrieval,chen2024benchmarking}. 
However, deploying RAG over privacy-sensitive corpora introduces substantial risks of knowledge-base exposure~\cite{zeng2024good,qi2024follow,peng2024data,he2025mitigating,zeng2025mitigating,he2025transform}. In practical pipelines, raw documents are often transformed into vectors through cloud-based or proprietary embedding services~\cite{enevoldsen2025mmteb} and stored in retrieval infrastructures such as vector databases, placing document embedding, indexing, and retrieval outside the full control of data owners. 
As a result, sensitive information may be exposed even before generation occurs. 
Moreover, attackers can exploit LLMs' instruction-following ability to induce RAG systems to reveal private content from retrieved documents~\cite{qi2024follow,zeng2024good}. 
Since post-retrieval defenses such as reranking or filtering cannot fully eliminate privacy risks once raw documents enter the model context~\cite{chase2022langchain,zeng2024good}, data-level protection before indexing becomes necessary.

\subsection{Data-Level Privacy Protection in RAG}

Classical privacy models, including $k$-anonymity, $l$-diversity, and $t$-closeness, provide important foundations for privacy protection~\cite{sweeney2002k,machanavajjhala2007diversity,li2006t}. 
However, these models are mainly designed for structured tabular data with explicit quasi-identifiers and sensitive attributes, whereas privacy-sensitive knowledge in RAG often appears in free-text documents with unstructured and context-dependent forms. 
As a result, free-text anonymization is commonly formulated as a detect-and-transform pipeline, where sensitive entities are first identified and then masked, replaced, generalized, or synthesized~\cite{dernoncourt2017identification,akbik2018contextual,devlin2019bert,stubbs2015annotating,deusser2025survey}. 
Building on this paradigm, existing data-level protection methods for RAG can be broadly categorized into synthesis-based and perturbation-based approaches. 
Synthesis-based methods reduce exposure to the original corpus by generating alternative texts that approximate private documents; for example, \citet{tang2023privacy} generated private in-context demonstrations from private datasets, while \citet{zeng2025mitigating} synthesized replacement documents for RAG knowledge bases using limited contextual information and adversarial privacy evaluation. 
Although such methods avoid directly storing raw private documents, they usually require extensive LLM-based generation and may introduce information loss, factual drift, or hallucinated content, thereby undermining the faithfulness and retrievability required by RAG. 
Perturbation-based methods instead protect text by modifying sensitive information or its representations. 
Differential Privacy (DP) provides a rigorous framework for limiting privacy leakage through randomized mechanisms~\cite{dwork2006calibrating,dwork2014algorithmic}, and has been applied to text at the token, sentence, and representation levels~\cite{feyisetan2020privacy,igamberdiev2023dp,plant2021cape}. 
For RAG knowledge bases, LPRAG~\cite{he2025mitigating} applies local differential privacy to sensitive entities detected in knowledge-base texts. 
However, DP-based perturbation may produce semantically imprecise or factually inconsistent replacements, which is problematic for RAG systems that rely on factual accuracy, entity consistency, and retrieval discriminability.

\textbf{Overall, existing data-level protection methods usually treat detected sensitive entities in a uniform manner, ignoring their heterogeneous roles in privacy leakage and RAG utility.} 
In practice, anonymizing a small subset of contextually critical entities may already be sufficient to break semantic inference paths, while further anonymization may unnecessarily remove topical cues, entity relations, and semantic structure. 
Therefore, the key challenge in data-level privacy protection for RAG is not simply how to anonymize sensitive entities, but how to selectively determine which entities should be anonymized under the privacy-utility trade-off. 
This motivates context-aware and utility-preserving entity anonymization for RAG knowledge bases.

\section{Problem Formulation}
\subsection{Problem Definition}
\label{sc:3.1}

In this section, we formalize the utility-preserving semantic privacy protection problem for RAG knowledge bases.

\begin{definition}[Knowledge-base Text Anonymization]
Given a \\
knowledge-base text \(T\), let 
\(\mathcal{E}_T=\{e_1,e_2,\ldots,e_n\}\) denote the set of extractable entities in \(T\). 
Let \(\Gamma\) be an entity-level generalization mapping that maps an entity to a coarse-grained semantic descriptor, and let \(\Pi_{\Gamma}\) denote the text-level anonymization mechanism induced by \(\Gamma\). 
For a selected entity subset \(\mathcal{G}_T\subseteq \mathcal{E}_T\), the anonymized text is defined as
\[
T^\prime=\Pi_{\Gamma}(T;\mathcal{G}_T),
\]
where each entity \(e\in\mathcal{G}_T\) is replaced with its generalized form \(\Gamma(e)\), while entities in \(\mathcal{E}_T\setminus\mathcal{G}_T\) remain unchanged.
\end{definition}

Existing privacy protection methods usually determine \(\mathcal{E}_T\) according to predefined sensitive categories. 
They often treats entities independently and often anonymizes all entities within protected categories. 
However, privacy leakage in knowledge-base texts is often contextual: whether a sentence reveals private information depends not only on individual entities but also on the contextual relationships among textual elements. 
Therefore, not all entities in \(\mathcal{E}_T\) necessarily require anonymization. Since privacy leakage is contextual and should be evaluated at the sentence level, anonymizing an appropriate subset \(\mathcal{G}_T\) of \(\mathcal{E}_T\) may already be sufficient to prevent privacy disclosure of the entire sentence.

\begin{definition}[Utility-Aware Privacy Protection]
Given a text \(T\) from knowledge base, entity set \(\mathcal{E}_T\),  and anonymization mechanism \(\Pi_{\Gamma}\), the task aims to select an anonymization subset \(\mathcal{G}_{T}\subseteq\mathcal{E}_T\) such that the anonymized text \(T^\prime=\Pi_{\Gamma}(T;\mathcal{G}_{T})\) satisfies the required privacy constraint while preserving as much task-relevant utility as possible.
\end{definition}

Let \(u(e)\ge 0\) denote the utility contribution of entity \(e\), and let \(R(T^\prime)\in[0,1]\) denote the residual semantic privacy risk of \(T^\prime\), i.e., the risk that private target entities in \(\mathcal{G}_{T}\) can still be reconstructed from the remaining context. 
The utility of the anonymized text is defined as the total utility of preserved entities:
\begin{equation}
    \mathcal{U}\big(T,\Pi_{\Gamma}(T;\mathcal{G}_{T})\big)
    =
    \sum_{e\in \mathcal{E}_T\setminus \mathcal{G}_{T}} u(e).
\end{equation}

Given an acceptable privacy-risk threshold \(\epsilon\), utility-preserving semantic privacy protection can be formulated as the following constrained combinatorial optimization problem:
\begin{equation}
\label{eq:optimization}
\begin{aligned}
\mathcal{G}_{T}^{*}
=
\underset{\mathcal{G}_{T}\subseteq \mathcal{E}_T}{\arg\max}
&\quad
\mathcal{U}\big(T,\Pi_{\Gamma}(T;\mathcal{G}_{T})\big) \\
\mathrm{s.t.}
&\quad
R\big(\Pi_{\Gamma}(T;\mathcal{G}_{T})\big)\le \epsilon .
\end{aligned}
\end{equation}

This formulation reflects the trade-off considered in our entity anonymization setting: increasing anonymization may help reduce residual privacy risk, but can also affect information useful for retrieval and generation.
Since finding the optimal subset \(\mathcal{G}_{T}^{*}\) requires searching over all possible entity subsets, the problem is combinatorial in nature and NP-hard, as shown in Appendix~\ref{ap:1}. 
TRIP-RAG therefore adopts a context-aware greedy approximation strategy, as detailed in Section~\ref{sc:4}.

\subsection{Threat Model}
\label{sc:3.2}

We define the threat model to specify the security meaning of the residual privacy risk \(R(T^\prime)\) in Eq.~\ref{eq:optimization}. 
We consider a polynomial-time attacker \(\mathcal{A}\) whose goal is to recover private target entities in \(\mathcal{G}_{T}\) from the anonymized text \(T^\prime=\Pi_{\Gamma}(T;\mathcal{G}_{T})\) and its remaining context. 
An attack succeeds if \(\mathcal{A}\) can reliably infer any protected entity. 
We consider two scenarios: in the \textit{untrusted-server scenario}, \(\mathcal{A}\) can directly inspect the stored anonymized texts; in the \textit{trusted-server scenario}, \(\mathcal{A}\) can only query the RAG system and observe its outputs. 
In both cases, \(\mathcal{A}\) is assumed to know the anonymization mechanism \(\Pi_{\Gamma}\), following the standard principle that privacy should not rely on mechanism secrecy. 
Under this threat model, \(R(T^\prime)\) denotes the residual contextual reconstruction risk of \(T^\prime\). 
We consider texts that violate the desired privacy requirement before anonymization, i.e.,\(R(T)>\epsilon\).
Thus, the privacy constraint in Eq.~\ref{eq:optimization} requires the anonymized text to reduce contextual reconstruction risk below the acceptable threshold \(\epsilon\).
To establish this property, we consider a relaxed adversarial objective and analyze the attacker's ability to distinguish between candidate original entities conditioned on the anonymized text. Intuitively, successful reconstruction implies successful distinguishing; therefore, bounding the distinguishing advantage provides a sufficient condition for limiting contextual reconstruction risk.
\begin{figure*} 
\centering 
\includegraphics[width=1\linewidth]{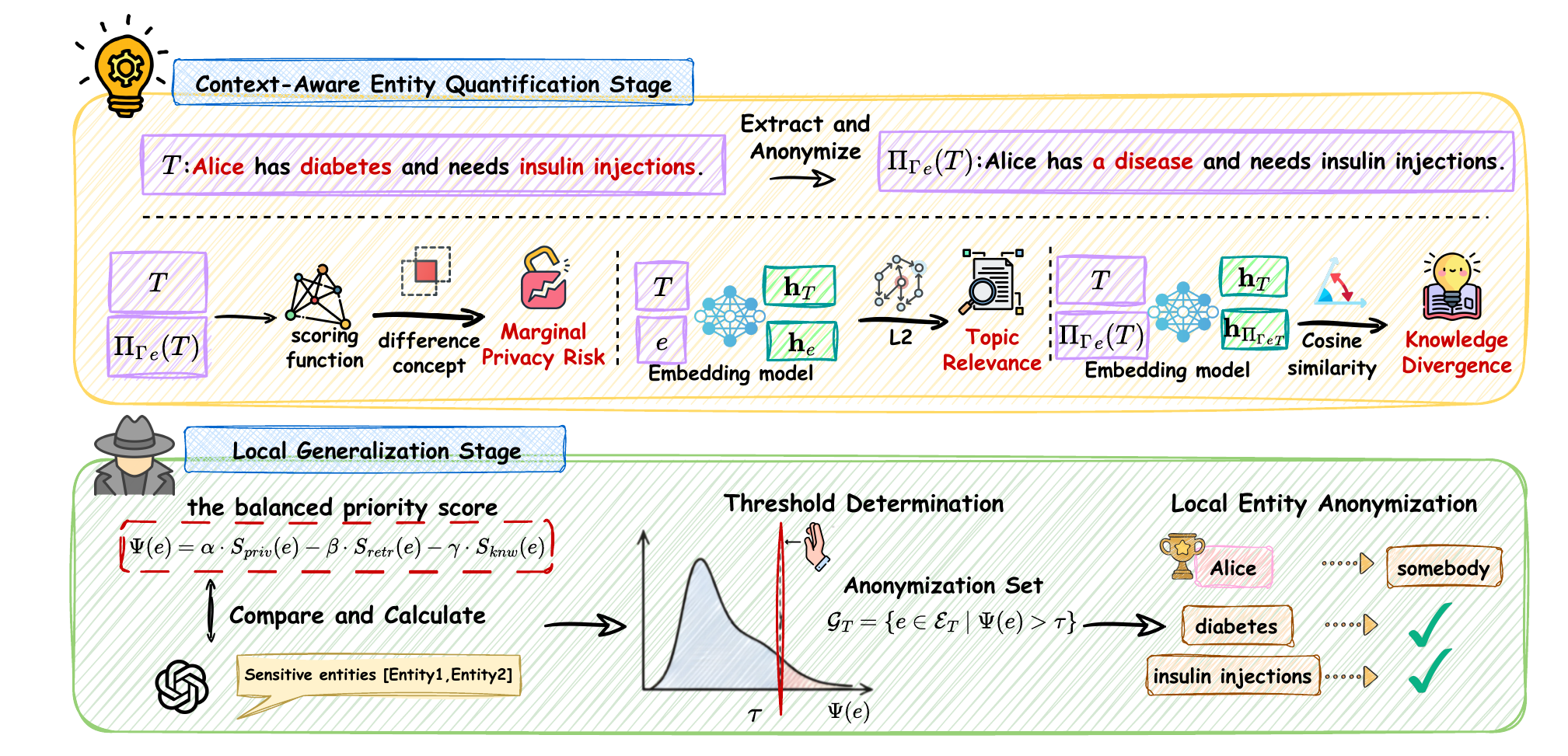} 
\caption{Overview of the TRIP-RAG framework. TRIP-RAG first extracts privacy-sensitive entities and evaluates their marginal privacy risk, topic relevance, and knowledge divergence. It then aggregates these dimensions into a priority score and applies a dataset-adaptive threshold to select entities for local anonymization.}
\label{fig:2} 
\label{fig:2}
\end{figure*}
\section{Methods}
\label{sc:4}
\subsection{Design and Workflow of TRIP-RAG}

To overcome the limitations of the ``one-size-fits-all'' text perturbation paradigm, we formalize the anonymization mechanism for RAG knowledge bases as a constrained combinatorial optimization problem. 
\textbf{The core intuition of TRIP-RAG is that different entities within a document possess heterogeneous impacts on semantic privacy risks and downstream task utility.}
As illustrated in Figure \ref{fig:2}, our framework addresses this problem through a two-stage pipeline.
First, the context-aware entity quantification stage evaluates the marginal contribution of each entity from both privacy and utility perspectives.
Second, the local generalization stage uses a greedy strategy to selectively anonymize high-priority entities, reducing privacy risks while preserving the document's semantic structure and retrieval utility.

\subsubsection{Context-Aware Entity Quantification Stage} 
Let $\mathcal{T}$ denote the high-dimensional textual semantic space of texts. 
Given an original document $T \in \mathcal{T}$ from an external knowledge base, we first deploy a bidirectional Transformer-based Named Entity Recognizer (NER), denoted as $\mathcal{F}_{NER}$, to extract the set of key entities \(\mathcal{E}_T=\{e_1,e_2,\ldots,e_n\}\)~\cite{zaratiana2024gliner}.
\textbf{Our objective is not to uniformly perturb the extractable entities $\mathcal{E}_T$, but to quantify the privacy-utility trade-off for each entity $e \in \mathcal{E}_T$.} 
We evaluate each entity along three dimensions: marginal privacy risk, knowledge divergence, and topical relevance. 
Experimental results show that the information overlap rate of these three scores is very low. 
Detailed analysis is provided in Appendix \ref{ap:8}.

\textbf{Marginal Privacy Risk ($S_{\mathrm{priv}}$).} 
Text privacy is highly context-dependent, making isolated entity evaluation suboptimal. 
To capture the contextual privacy leakage induced by a specific entity, we use a local anonymization operator ${\Pi_{\Gamma}}_{e}(T)$ defined in Section \ref{sc:3.1} to mask the entity $e$ within the text $T$. 
Let $f_{\mathrm{priv}}: \mathcal{T} \rightarrow \mathbb{R}$ be a privacy-aware surrogate function parameterized by a pre-trained language model, which maps a document to its absolute privacy risk score and is empirically calibrated using GDPR guidelines and validated against human judgements through correlation analysis; see Appendix \ref{ap:5}). 
We define the marginal privacy risk of entity $e$ as the reduction in document-level privacy risk after locally anonymizing $e$:
\begin{equation}
	S_{\mathrm{priv}}(e)=f_{\mathrm{priv}}(T)-f_{\mathrm{priv}}({\Pi_{\Gamma}}_{e}(T)),
\end{equation}
This differential formulation enables scalable entity-level risk estimation while accounting for the surrounding textual context.

\textbf{Knowledge Divergence ($S_{\mathrm{knw}}$).} 
While privacy protection requires removing or generalizing risky entities, RAG knowledge bases must also preserve the semantic structure of the document to support reliable retrieval and generation. 
Entities that strongly affect the document representation often carry important factual or relational knowledge; anonymizing them may therefore degrade knowledge utility. 
Based on the semantic clustering properties of distributed representations~\cite{mikolov2013efficient}, we measure how much the document representation changes when a specific entity is anonymized.
Let \(\Phi: \mathcal{T}\rightarrow \mathbb{R}^{d}\) denote a pre-trained text encoder that maps an input text into a \(d\)-dimensional semantic representation. 
For a document $T$ and its perturbed version ${\Pi_{\Gamma}}_{e}(T)$, let $\mathbf{h}_T=\Phi(T)$ and $\mathbf{h}_{{\Pi_{\Gamma}}_{e}(T)}=\Phi({\Pi_{\Gamma}}_{e}(T))$ denote their respective latent representations. 
We quantify the knowledge divergence by evaluating the cosine distance between them:
\begin{equation}		
	\resizebox{0.8\linewidth}{!}{$
		S_{\mathrm{knw}}(e)=\mathcal{F}_{\mathrm{norm}}\left(1-\frac{\Phi(T)\cdot\Phi({\Pi_{\Gamma}}_{e}(T))}{||\Phi(T)||_2||\Phi({\Pi_{\Gamma}}_{e}(T))||_2}\right),
		$}		
\end{equation} 
where $\mathcal{F}_{\mathrm{norm}}(\cdot)$ represents a robust normalization function. A larger angular shift signifies a higher contribution of the entity to the overall knowledge structure.

\textbf{Topical Relevance ($S_{retr}$).} RAG systems inherently rely on vector space geometry to conduct similarity search ~\cite{lewis2020retrieval}. For the document $T$ and the entity $e$, let $\mathbf{h}_T=\mathcal{E}(T)$ and $\mathbf{h}_e=\mathcal{E}(e)$ denote their latent representations, respectively. We utilize the entity to simulate a retrieval query, thereby measuring the $L_2$ divergence between the entity topic and the global document topic to quantify the topic relevance of $e$ to the document:
\begin{equation}	
	S_{\mathrm{retr}}(e)=\mathcal{F}_\mathrm{{norm}}(-||\mathbf{h}_e-\mathbf{h}_T||_2),	
\end{equation} 

A smaller divergence indicates that the entity is more relevant to the retrieval of the document.
\subsubsection{Local Generalization Stage} 
\label{sc:4.1.2}
In the second stage, the framework identifies the optimal subset of entities to be anonymized by solving a surrogate optimization objective. We aggregate the three decoupled dimensions mentioned above into a unified entity priority score $\Psi(e)$, parameterized by weighting coefficients $\alpha, \beta, \gamma \ge 0$:
\begin{equation}
	\Psi(e)=\alpha S_\mathrm{{priv}}(e)-\beta S_\mathrm{retr}(e)-\gamma S_\mathrm{knw}(e),
\end{equation}This formulation balances the objective of maximizing the reduction of privacy risks (driven by $S_\mathrm{priv}$) against the penalties of utility degradation (constrained by $S_\mathrm{retr}$ and $S_\mathrm{knw}$).

\textbf{Threshold Determination.} Given that different vertical domains (e.g., healthcare and financial services) exhibit extreme heterogeneity in privacy sensitivity distributions, traditional static Top-K truncation strategies are highly susceptible to destroying the topological structure of the high-dimensional semantic space, thus leading to suboptimal solutions with over-anonymization. Recalling the problem formulation in Section \ref{sc:3.1}, since semantic privacy possesses strong context dependence, we introduce a Large Language Model (LLM) as an evaluator. By executing a single-turn dialogue on a small number of samples, we can extract the critical subset of entities necessary to eliminate semantic privacy risks. By comparing the $\Psi(e)$ scores of these entities, we can empirically determine a safe critical threshold $\tau$. Specifically, we bound the subset $\mathcal{G}_T$ of entities to be anonymized based on the obtained empirical threshold, incorporating entities with scores exceeding the threshold into this set. We additionally conducted a manual secondary evaluation of the LLM's assessment results; detailed settings can be found in Appendix \ref{ap:6}:
\begin{equation}
	\mathcal{G}_T = \{e \in \mathcal{E}_T \mid \Psi(e) > \tau \},
\end{equation}

\textbf{Semantics-Preserving Entity Anonymization.} To ensure that the anonymized document $T^\prime$ can maintain the strict syntactic coherence and factual consistency required for downstream LLM reasoning, we define a secure generalization mapping $\mathcal{M}: \mathcal{L} \rightarrow \mathcal{D}$, which projects the original semantic label $\ell_e \in \mathcal{L}$ of entity $e$ into a safe, coarse-grained descriptor space $\mathcal{D}$.
Detailed replacement examples are presented in the case study in Section~\ref{sc:case}. The system iteratively applies the transformation $\Gamma(e)=\mathcal{M}(\ell_e)$ to all $e \in \mathcal{G}_T$. At this point, TRIP-RAG successfully neutralizes exact sensitive attributes while strictly preserving the structure and semantic topology of the external knowledge base.
\subsection{Security Analysis}
Given the attacker's objectives outlined in Section \ref{sc:3.2}, we need to analyze the semantic security of the proposed method. 
\subsubsection{Proof of Semantic Security} 
If, after observing the anonymized text $T^\prime={\Pi_{\Gamma}}(T)$, the advantage of an attacker in distinguishing between any two original entities $e_0,e_1$ belonging to the same generalization class $C$ is negligible, i.e.,
\begin{equation}
	\resizebox{0.85\linewidth}{!}{$
		Adv_{\mathcal{A}}^{sem}\left(\lambda\right)=\left|\Pr\left[\mathcal{A}\left({\Pi_{\Gamma}}\left(T_{e_0}\right)\right)=1\right]
		-\Pr\left[\mathcal{A}\left({\Pi_{\Gamma}}\left(T_{e_1}\right)\right)=1\right]\right|\le\epsilon\left(\lambda\right),
		$}
\end{equation}
then the mechanism $\Pi$ is said to satisfy semantic security. It is easy to see that the generalization function $\Gamma$ is a non-injective surjection. Furthermore, since the objects being generalized, such as names, are highly extensive, the pre-image cardinality $\left|\Gamma^{-1}\left(c\right)\right| $ of the generalization class is sufficiently large. \textbf{Therefore, the entity generalization mechanism of TRIP-RAG achieves semantic indistinguishability against an attacker in polynomial time, i.e., $Adv_{\mathcal{A}}^{sem}\approx0$.} The detailed proof can be found in Appendix \ref{ap:2}. 
\begin{table*}[htbp]		
    \caption{Utility evaluation results. Origin represents the upper-bound performance without anonymization. Bold values indicate the best performance among privacy-preserving methods under each model. The blue downward percentages denote the relative drop from Origin, where smaller drops indicate better utility preservation.}
    \vspace{-0.1in}
	\label{tb:2}		
	\centering		
	\definecolor{origingray}{HTML}{F2F2F2}		
	\definecolor{tripblue}{HTML}{E6F0FA}		
	\definecolor{dropgray}{HTML}{2E5077}
	\newcommand{\drop}[1]{\ensuremath{{\color{dropgray}\downarrow#1\%}}}		
	\renewcommand{\arraystretch}{1.2} 
	\resizebox{\textwidth}{!}{		
		\begin{tabular}{llcccccccccc}			
			\toprule			
			\multirow{2}{*}{\large Model} & \multirow{2}{*}{\large Method} & \multicolumn{5}{c}{\large ChatDoctor } & \multicolumn{5}{c}{\large EnronQA} \\				
			\cmidrule(lr){3-7} \cmidrule(lr){8-12}			
			& & \large BLEU & \large ROUGE-L & \large Recall@1 & \large Recall@5 & \large Recall@10 & \large BLEU & \large ROUGE-L & \large Recall@1 & \large Recall@5 & \large Recall@10\\			
			\midrule			
			\rowcolor{origingray} 			
			\cellcolor{white} & Origin& 0.156 & 0.132 & 1.000 & 1.000 & 1.000 & 0.559 & 0.691 & 1.000 & 1.000 & 1.000 \\			
			& Redact& 0.078\drop{50} & 0.078\drop{41} & 0.430\drop{57} & 0.524\drop{48} & 0.568\drop{43} & 0.230\drop{59} & 0.269\drop{61} & 0.782\drop{22} & 0.455\drop{54} & 0.400\drop{60} \\		
			& Paraphrase & 0.060\drop{62} & 0.066\drop{50} & 0.161\drop{84} & 0.235\drop{76} & 0.266\drop{73} & 0.083\drop{85} & 0.121\drop{82} & 0.661\drop{34} & 0.497\drop{50} & 0.466\drop{53} \\			
			& LPRAG& 0.070\drop{55} & 0.075\drop{43} & 0.389\drop{61} & 0.477\drop{52} & 0.513\drop{49} & 0.280\drop{50} & 0.326\drop{53} & 0.847\drop{15} & 0.597\drop{40} & 0.559\drop{44} \\			
			\rowcolor{tripblue} 			
			\multirow{-5}{*}{\cellcolor{white}Qwen3-0.6B} 		
			& TRIP-RAG& \textbf{0.083}\drop{47} & \textbf{0.084}\drop{36} & \textbf{0.721}\drop{28} & \textbf{0.784}\drop{22} & \textbf{0.810}\drop{19} & \textbf{0.364}\drop{35} & \textbf{0.446}\drop{35} & \textbf{0.935}\drop{6} & \textbf{0.736}\drop{26} & \textbf{0.710}\drop{29} \\			
			\midrule		
			\rowcolor{origingray} 		
			\cellcolor{white} & Origin& 0.162 & 0.136 & 1.000 & 1.000 & 1.000 & 0.563 & 0.694 & 1.000 & 1.000 & 1.000 \\		
			& Redact& 0.072\drop{56} & 0.075\drop{45} & 0.415\drop{58} & 0.514\drop{49} & 0.547\drop{45} & 0.237\drop{58} & 0.276\drop{60} & 0.743\drop{26} & 0.411\drop{59} & 0.363\drop{64} \\		
			& Paraphrase & 0.063\drop{61} & 0.067\drop{51} & 0.130\drop{87} & 0.188\drop{81} & 0.211\drop{79} & 0.076\drop{87} & 0.112\drop{84} & 0.599\drop{40} & 0.397\drop{60} & 0.348\drop{65} \\	
			& LPRAG& 0.066\drop{59} & 0.072\drop{47} & 0.346\drop{65} & 0.446\drop{55} & 0.478\drop{52} & 0.295\drop{48} & 0.342\drop{51} & 0.821\drop{18} & 0.549\drop{45} & 0.498\drop{50} \\		
			\rowcolor{tripblue} 		
			\multirow{-5}{*}{\cellcolor{white}bge-large-en-v1.5} 
			& TRIP-RAG& \textbf{0.086}\drop{47} & \textbf{0.088}\drop{35} & \textbf{0.672}\drop{33} & \textbf{0.752}\drop{25} & \textbf{0.778}\drop{22} & \textbf{0.393}\drop{30} & \textbf{0.477}\drop{31} & \textbf{0.923}\drop{8} & \textbf{0.695}\drop{31} & \textbf{0.659}\drop{34} \\		
			\midrule
			\rowcolor{origingray} 		
			\cellcolor{white} & Origin& 0.160 & 0.135 & 1.000 & 1.000 & 1.000 & 0.568 & 0.698 & 1.000 & 1.000 & 1.000 \\		
			& Redact& 0.067\drop{58} & 0.074\drop{45} & 0.416\drop{58} & 0.512\drop{49} & 0.545\drop{45} & 0.213\drop{63} & 0.249\drop{64} & 0.747\drop{25} & 0.447\drop{55} & 0.406\drop{59} \\		
			& Paraphrase & 0.059\drop{63} & 0.066\drop{51} & 0.105\drop{90} & 0.180\drop{82} & 0.205\drop{80} & 0.075\drop{87} & 0.110\drop{84} & 0.598\drop{40} & 0.388\drop{61} & 0.351\drop{65} \\	
			& LPRAG& 0.057\drop{64} & 0.065\drop{52} & 0.360\drop{64} & 0.443\drop{56} & 0.480\drop{52} & 0.258\drop{55} & 0.299\drop{57} & 0.776\drop{22} & 0.496\drop{50} & 0.452\drop{55} \\
			\rowcolor{tripblue} 		
			\multirow{-5}{*}{\cellcolor{white}all-MiniLM-L6-v2} 
			& TRIP-RAG& \textbf{0.096}\drop{40} & \textbf{0.095}\drop{30} & \textbf{0.611}\drop{40} & \textbf{0.683}\drop{32} & \textbf{0.731}\drop{27} & \textbf{0.364}\drop{36} & \textbf{0.444}\drop{36} & \textbf{0.929}\drop{7} & \textbf{0.702}\drop{30} & \textbf{0.679}\drop{32} \\
			\bottomrule
		\end{tabular}
	}
\end{table*}

\section{Experiment}
We evaluate TRIP-RAG through cross-dataset and multi-model experiments to address the following research questions (RQ):
\begin{rqbox}
\textbf{RQ1: } How does TRIP-RAG balance privacy protection and task utility compared to state-of-the-art baselines in privacy-sensitive knowledge-intensive tasks?

\textbf{RQ2: } How do Knowledge Divergence and Topical Relevance affect task performance relative to Marginal Privacy Risk alone, and how can their associated hyperparameters be optimized?

\textbf{RQ3: } How does sample size influence the alignment of Marginal Privacy Risk with human judgment?
\end{rqbox}

\subsection{Experimental Settings} 
\label{sc:5.1}
\subsubsection{Datasets and Evaluation Metrics}
We evaluate our method on two privacy-sensitive RAG scenarios: healthcare consultation and workplace communication. For the healthcare scenario, we adopt ChatDoctor-HealthCareMagic-100k~\cite{he2025mitigating}, which contains over 100,000 doctor-patient conversations involving personal information such as names, ages, and health-related descriptions. For the workplace communication scenario, we use EnronQA~\cite{ryan2025enronqa}, which is constructed from the Enron email corpus and contains 70,000 real-world workplace question-answer pairs with private information such as personal names and phone numbers.
Following prior work, we use 99\% of each dataset as the retrieval corpus and the remaining 1\% as the test set. We evaluate each method from three task-oriented perspectives. First, to assess whether anonymization preserves the retrievability of useful knowledge, we evaluate retrieval performance using Recall@k, following \citet{thakur2021beir}. Second, to measure whether the anonymized knowledge still preserves useful textual information, we report ROUGE-L~\cite{lin2004rouge} and BLEU~\cite{papineni2002bleu} between the original text and the anonymized or reconstructed text. Third, to evaluate privacy protection, we follow \citet{he2025mitigating} and adopt the privacy extraction attack proposed by \citet{zeng2024good}, which measures how many private entities can be inferred or extracted from the LLM-generated responses.

\subsubsection{Baselines and Implementations}
To verify the effectiveness of our method, we incorporated three baselines, including full generalized mapping replacement, paraphrasing methods~\cite{jainbaseline2023}, and a representative existing knowledge base anonymization method—namely, the anonymization method based on local differential privacy proposed by \citet{he2025mitigating}. Detailed information regarding these methods is provided in Appendix \ref{ap:4}. Additionally, we included the results from the original data for comparison.Both the storage of embeddings and the construction of the retrieval database were implemented using the FAISS library. By default, we employed $L_2$-norm as the distance metric to compare embeddings. In experiments outside the scope of exploring retrieval performance, we uniformly retrieved 5 documents to serve as the Context. For entity extraction and the batch scoring of Marginal Privacy Risk, we utilized the gliner-v2.1 model ~\cite{zaratiana2024gliner} and the deberta-v3-base model ~\cite{he2021debertav3}, respectively. For $\alpha$, $\beta$, and $\gamma$, we assigned the values of 1, 0.5, and 0.4, respectively, and following the threshold determination mechanism outlined in Section \ref{sc:4.1.2}, we established the threshold for ChatDoctor-HealthCareMagic-100k at 0.0526, and the threshold for Enronqa at 0.2675 . All experimental evaluations were executed on an NVIDIA RTX 4090 GPU.


\begin{table*}[htbp]
	\centering
	\caption{Example of comparison of anonymization methods}
	\label{tb:5}
	\renewcommand{\arraystretch}{1.1} 
    \resizebox{0.85\linewidth}{!}{
	\begin{tabularx}{\linewidth}{p{0.1\linewidth}X} 
			\toprule
			\textbf{Method} & \textbf{Example} \\
			\midrule
			\textbf{Origin} & 
			Patient: Hello, I am \originhl{32 year old} \originhl{male},3 months back I have married but still no hope for baby. Is taking \originhl{Coq forte} /\originhl{ubiq300}/\originhl{siotone} a good option for conceive a child?\newline Doctor: Ask her to start \originhl{folic acid tablets}, healthy balanced diet and moderate exercise. \\
			\midrule
			\textbf{TRIP-RAG} & 
			Patient: Hello, I am \triphl{a certain age} \triphl{a gender}, 3 months back I have married but still no hope for baby. Is taking \triphl{Coq forte} /\triphl{ubiq300}/\triphl{siotone} a good option for conceive a child?\newline Doctor: Ask her to start \triphl{folic acid tablets}, healthy balanced diet and moderate exercise. \\
			\midrule
			\textbf{LPRAG} & 
			Patient: Hello, I am \lpraghl{29 Mathilde Stadt} \lpraghl{male}, 3 months back I have married but still no hope for baby. Is taking Same \lpraghl{Wanting} /\lpraghl{Wanting}/\lpraghl{Allegedly} a good option for conceive a child?\newline Doctor:  Ask her to start \lpraghl{Witness ̫ \textbackslash u305D\textbackslash u306E}, healthy balanced diet and moderate exercise. \\
			\midrule
			\textbf{Paraphrase} & 
			1. Patient's age: \parahl{32 years old}.\newline  2. Doctor's recommendation: Ask the wife to start \parahl{folic acid tablets}, healthy balanced diet, and moderate exercise. \\		
			\midrule	
			\textbf{Redact} & 		
			Patient: Hello,I am \redacthl{a certain age} \redacthl{a gender}, 3 months back I have married but still no hope for baby. Is taking \redacthl{a medication} /\redacthl{a medication}/\redacthl{a medication} a good option for conceive a child?\newline Doctor: Ask her to start \redacthl{a medication}, healthy balanced diet and moderate exercise. \\		
			\bottomrule		
		\end{tabularx}%
        }
\end{table*}

\subsection{Main Results(RQ1)}
\label{sc:5.2}
We conducted extensive experiments across multiple datasets and models to evaluate retrieval performance, answer utility, and security. The experimental results further demonstrate the superiority and advancement of our framework.

\subsubsection{Utility Evaluation}
To evaluate the utility of the anonymized knowledge base, we report ROUGE-L and BLEU between the generated answers and the ground-truth answers to measure generation quality, and use Recall@k to evaluate retrieval consistency by comparing the overlap between the documents retrieved from the original and anonymized knowledge bases under the same queries. As shown in Table~\ref{tb:2}, TRIP-RAG consistently preserves stronger task utility than the baseline methods across both retrieval and generation metrics. Compared with rewriting-based methods such as Paraphrase, TRIP-RAG better maintains the original knowledge structure and achieves the best Recall@k scores, improving over the baselines by 9\%--60\%. This suggests that aggressive rewriting may alter the retrievable semantics of the original documents, thus weakening retrieval consistency after anonymization. Compared with full-anonymization methods such as LPRAG, TRIP-RAG also obtains the best BLEU and ROUGE-L scores, with improvements of 3\%--56\% and 5\%--53\%, respectively. These results indicate that treating all sensitive entities uniformly may remove information useful for answer generation, whereas our context-aware entity selection strategy better preserves task-relevant content while performing anonymization.


\subsubsection{Privacy Protection Analysis}

\begin{table}
	\centering
	\caption{Privacy leakage rates under the privacy extraction attack.}
	\label{tb:3}
	\definecolor{origingray}{HTML}{F2F2F2}
	\definecolor{tripfen}{HTML}{FFF0F5}
	\renewcommand{\arraystretch}{1.25} 	
	\small
	\begin{tabular}{lcc}			
		\toprule		
		\multicolumn{1}{c}{Method} & ChatDoctor & Enronqa \\			
		\midrule			
		\rowcolor{origingray} 		
		\cellcolor{white}Origin & 0.23 & 0.63 \\		
		Paraphrase& \underline{0.19} & \textbf{0.26} \\			
		LPRAG& 0.21 & 0.48 \\		
		\rowcolor{tripfen} 			
		\cellcolor{white}TRIP-RAG & \textbf{0.17} & \underline{0.33} \\		
		\bottomrule
	\end{tabular}
\end{table}


We evaluate privacy protection using the privacy extraction attack proposed by \citet{zeng2024good}. In this setting, an anonymization method is considered more effective if the RAG system exposes fewer sensitive entities when answering questions, since such exposure indicates contextual re-identification from the retrieved content. As reported in Table~\ref{tb:3}, TRIP-RAG consistently reduces privacy leakage on both datasets, achieving a 26\% reduction on ChatDoctor and a 48\% reduction on EnronQA compared with the original data. These results demonstrate that TRIP-RAG can effectively suppress privacy exposure in privacy-sensitive RAG scenarios. While Paraphrase obtains slightly lower leakage on EnronQA, it significantly compromises utility, as reflected by the results in Table~\ref{tb:2}. Therefore, its privacy gain is largely achieved through excessive information removal. By contrast, TRIP-RAG maintains strong utility while substantially reducing privacy leakage, demonstrating its effectiveness in balancing privacy protection and downstream task performance.
\begin{figure*}[htbp]
	\centering
	\includegraphics[width=0.95\linewidth]{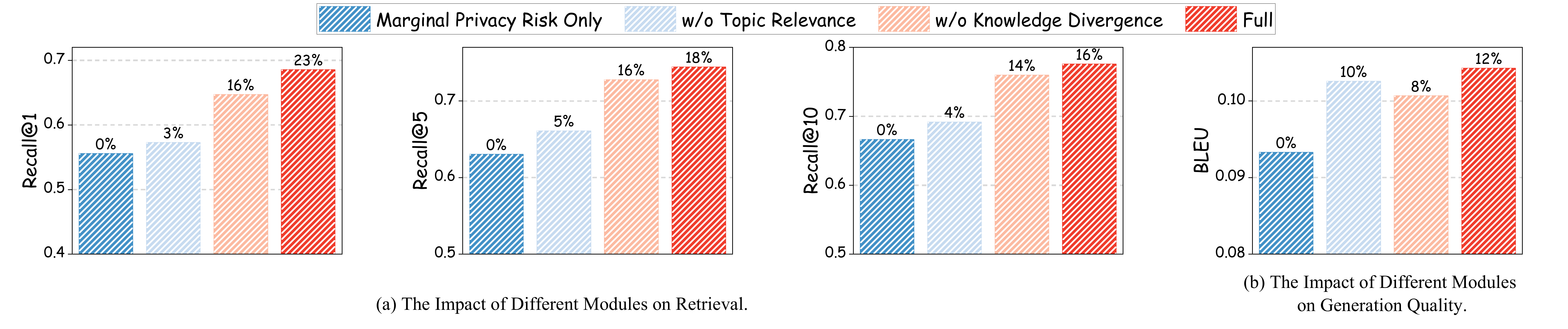} 
	\caption{Dimensional Ablation Study Evaluating the Contribution of Distinct Context-Aware Modules.}
	\label{fig:4}
\end{figure*}
\begin{figure*}[htbp]
	\centering
	\includegraphics[width=0.95\linewidth]{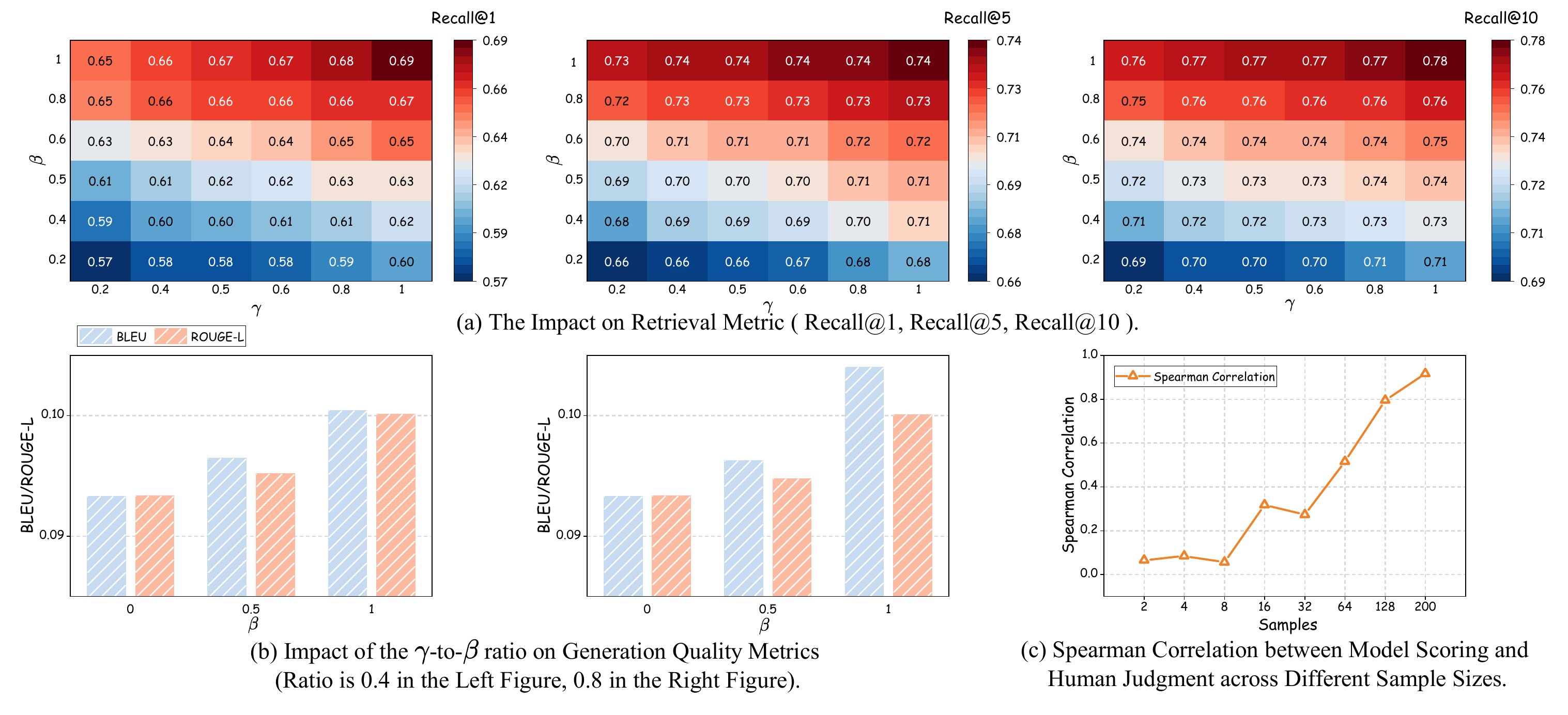} 
	\caption{Hyperparameter Sensitivity Analysis and the Impact of Labeled Data Size on Privacy Scoring.}
	\label{fig:5}
\end{figure*}
\subsection{Case Study(RQ1)}
\label{sc:case}
This section presents a case study illustrating how TRIP-RAG achieves privacy protection while preserving knowledge utility.

As shown in Table~\ref{tb:5}, the \origintexthl{original} consultation contains both demographic information and treatment-related medical knowledge. Existing anonymization methods either introduce semantic distortion (\lpragtexthl{LPRAG}), remove utility-critical entities through uniform masking (\redacttexthl{Redact}), or weaken the original interaction context (\paratexthl{Paraphrase}).

In contrast, \triptexthl{TRIP-RAG} selectively anonymizes entities according to contextual privacy contribution. Demographic attributes are generalized to eliminate identity-related inference signals, while medications and treatment recommendations are preserved as they remain essential for medical reasoning. In this way, TRIP-RAG transforms the original privacy-bearing consultation into a privacy-safe knowledge representation, eliminating sentence-level privacy risk while preserving retrieval and answer utility. This example demonstrates that local anonymization can achieve global semantic privacy protection without sacrificing utility-critical knowledge.
\begin{rqbox}
\textbf{Answer to RQ1: }TRIP-RAG achieves comparable privacy protection to state-of-the-art baselines while preserving substantially higher task utility.
\end{rqbox}

\subsection{Ablation Study(RQ2)} 
\label{sc:5.3}

We conducted an ablation study on TRIP-RAG to evaluate the rationality of the framework design. We utilized the all-MiniLM-L6-v2 as the embedding model and evaluated our method on the ChatDoctor dataset. 

Given the necessity of privacy protection, we investigate the advantages of incorporating topic relevance and knowledge divergence into the framework. As shown in Figure \ref{fig:4}, compared to the model based solely on marginal privacy risk, the inclusion of topical relevance yields improvements of 16\% in the retrieval metrics Recall@1. This indicates that some privacy entities also serve as important retrieval cues, and preserving them when appropriate can improve retrievability.Similarly, incorporating knowledge divergence leads to a 10\% improvement in the utility metric BLEU-1, showing that controlling semantic changes helps maintain answer quality. The full model achieves overall improvements of 23\%, 18\%, and 12\% in Recall@1, Recall@5, and BLEU-1, respectively, suggesting that the three dimensions provide complementary guidance for selecting entities to anonymize.

%

\subsection{Hyperparameter Sensitivity Analysis(RQ2)} 
As shown in Figure~\ref{fig:5}, we fix $\alpha=1$ and vary $\beta$ and $\gamma$ over \{0.2, 0.4, 0.5, 0.6, 0.8, 1.0\} to analyze the influence of topical relevance and knowledge divergence. Retrieval performance exhibits a consistent upward trend as $\beta$ increases, indicating that assigning greater weight to topical relevance helps preserve entities that contribute strongly to document representation and topic matching. Consequently, anonymized texts retain more retrieval-critical information and achieve higher Recall@k.However, generation quality depends not only on retrieval effectiveness but also on preserving semantic fidelity after anonymization. Increasing $\beta$ alone may retain retrieval cues while introducing unnecessary semantic shifts, whereas increasing $\gamma$ encourages the anonymized text to remain closer to the original knowledge representation. Therefore, we search for a balanced weighting between topical relevance and knowledge divergence to jointly optimize retrieval and answer quality. Based on the average BLEU score, we find that the best performance is achieved at $\gamma/\beta=0.8$, which is adopted as the default configuration in subsequent experiments.
\begin{rqbox}
\textbf{Answer to RQ2: }Topical relevance improves retrieval by preserving retrieval-critical entities, while knowledge divergence maintains answer quality through semantic fidelity; combining both yields the best overall utility at $\gamma/\beta=0.8$.
\end{rqbox}


\subsection{Impact of Labeled Data on Marginal Privacy Risk(RQ3)}
We selected sample sizes of 2, 4, 6, 8, 16, 32, 64, 128, and 200 for training, and calculated the Spearman correlation coefficient against manually annotated test samples. The results, as illustrated in the Figure \ref{fig:5}, demonstrate that as the number of samples increases, the model's scoring progressively aligns with human judgment standards. At a sample size of 128, a strong positive correlation of nearly 0.8 is achieved. With the 200 samples utilized in this study, the correlation coefficient reaches 0.9160, indicating that the model is fundamentally consistent with human judgment. 
\begin{rqbox}
\textbf{Answer to RQ3: }Marginal Privacy Risk shows progressively improved agreement with human judgment as training data increases and achieves strong alignment from 128 samples onward.
\end{rqbox}









\section{Conclusions}
We highlight the distinction between text semantic-level privacy protection and scope-entity privacy protection, correcting the misconceptions regarding text semantic privacy protection from the perspective of problem formulation. We focus on the variability of privacy protection intensity and the disparities among entities, constructing a context-aware anonymization framework tailored for knowledge databases by defining the marginal privacy risk, knowledge divergence, and topical relevance of entities. Through cross-dataset and multi-model experiments, we demonstrate that our proposed scheme quantifies and balances privacy and utility, achieving a percentage improvement in utility while having almost no impact on privacy. The evaluation framework proposed in this paper can also be extended to the broader field of text anonymization and integrated with various downstream anonymization methods, providing a novel reference and a controllable-intensity perspective for text semantic privacy protection.
\section{Limitation}
In this study, we rely on the GLiNER model to extract private entities ~\cite{zaratiana2024gliner} for the experimental portion of the paper. Although our investigation indicates that this custom-label-based extraction method is currently the state-of-the-art and the most suitable approach for privacy protection scenarios, our manual audit of the extraction results reveals that this method still misses some entities that appear in uncommon positions. Nevertheless, our overall experimental and evaluation results are based exclusively on the identified entity content, which does not compromise the experimental findings of this paper, but we look forward to future breakthroughs in entity extraction research.

\begin{acks}
To Robert, for the bagels and explaining CMYK and color spaces.
\end{acks}


\appendix
\section{Reduction of the Problem of Finding the Optimal Generalized Entity Subset}
\label{ap:1}
We prove this by performing a polynomial-time reduction from the classical 0-1 knapsack problem to our anonymization optimization problem. Consider a special case scenario where the background knowledge of the attacker faced by the system is scarce, such that the generalization of any single entity is sufficient to satisfy the anti-reconstruction entropy constraint $\eta$. Under this special case, non-reconstructability becomes a non-binding constraint and can be ignored. At this point, the optimization problem degenerates into: selecting a subset of entities $X$ subject to a total capacity limit, such that the total ``value'' $u_e$ of these entities is maximized. In this mapping: the entity's privacy $S_{priv}^e$ is equivalent to the weight of an item in the knapsack problem; the entity's utility contribution $u_e$ is equivalent to the value of an item in the knapsack problem; and the privacy threshold $B_{priv}$ is equivalent to the maximum capacity of the knapsack. Since the 0-1 knapsack problem is universally recognized as an NP-Hard problem in the classical combinatorial optimization field, and the aforementioned special case is merely a proper subset of the problem proposed in this paper, the original problem is at least as hard as the 0-1 knapsack problem. Namely, this problem is strongly NP-Hard.
\section{Information Overlap Among the Three Features}
\label{ap:8}
We employ the Spearman correlation coefficient to investigate whether there is an information overlap among the three features. As demonstrated in the Table \ref{tb:feature_overlap}, the maximum correlation coefficient is merely 0.4166. By calculating the square of the correlation coefficients to estimate the proportion of their shared rank variance, it can be observed that at least 80\% of the information remains independent. This proves the distinctiveness of our evaluation dimensions from the perspective of the information overlap rate.
\begin{table}[htbp]
	\centering
	\caption{Information Overlap Among the Three Features (Spearman $\rho$)}
	\label{tb:feature_overlap}
	\renewcommand{\arraystretch}{1.25} 
    \setlength{\tabcolsep}{3pt}
	\begin{tabular}{lcc}
		\toprule
		Feature Pair& ChatDoctor & Enronqa \\
		\midrule
		$S_{priv}$ vs $S_{knw}$ & +0.2070 & +0.0456 \\
		$S_{priv}$ vs $S_{retr}$ & +0.1342 & +0.0764 \\
		$S_{knw}$ vs $S_{retr}$ & +0.3923 & +0.4166 \\
		\bottomrule
	\end{tabular}
\end{table}
\section{Privacy Annotation Details}
\label{ap:5}
\subsection{Data Sampling and Annotation Guidelines}
To construct a high-quality training dataset for privacy scores, we adopted a random sampling method. We randomly sampled 200 representative text snippets rich in entities from the dataset, respectively. We recruited two researchers with backgrounds in Natural Language Processing (NLP) and foundational knowledge in data privacy protection to serve as independent annotators. Prior to formal annotation, all annotators underwent specialized training focused on the core privacy definitions of the General Data Protection Regulation (GDPR) to ensure consistency in annotation standards.

The specific regulation content we referred to is as follows:

\textit{‘personal data’ means any information relating to an identified or identifiable natural person (‘data subject’); an identifiable natural person is one who can be identified, directly or indirectly, in particular by reference to an identifier such as a name, an identification number, location data, an online identifier or to one or more factors specific to the physical, physiological, genetic, mental, economic, cultural or social identity of that natural person;}

We established the following 4-level scoring rubric to quantify the privacy importance score of entities:
\begin{itemize}
	\item Level 3 (Score 0.7-1.0): Entities that can directly and uniquely pinpoint a specific individual. Examples: real names, identification numbers, telephone numbers, exact home addresses, etc.
	\item Level 2 (Score 0.4-0.7): Entities posing moderate isolated risks but potential inference risks in context, or personal information susceptible to temporal factors.Examples: real age, biometric characteristics.
	\item Level 1 (Score 0.1-0.4): Entities that contain certain personal characteristics but possess a high degree of generalization within a large population, or entities for which auxiliary inference information is difficult to obtain. Examples: broad geographical regions (e.g., California), medical diagnostic results, and major business decisions.
	\item Level 0 (Score 0.0-0.1): General vocabulary or objective medical/business terminology that essentially does not involve personal privacy. Examples: common drug names (e.g., insulin), common symptoms, and general corporate operations.
\end{itemize}
\subsection{Comparison Between Language Model and Human Evaluation}
To validate the reliability of the language model's privacy risk evaluation, we conducted a correlation analysis between the model's assigned scores and the human annotations. We utilized the Spearman correlation coefficient to measure the monotonic relationship between the two sets of evaluations. The analysis demonstrated a highly strong alignment, achieving a Spearman correlation coefficient of $0.9139$ on the ChatDoctor dataset and $0.9085$ on the Enronqa dataset. These results indicate that the privacy-aware surrogate function's evaluations are fundamentally consistent with human judgment standards.
\section{Threshold Determination Details}
\label{ap:6}
We designed the following single-turn dialogue prompt for the LLM to guide it to focus on the combinatorial inference risks arising from intertwined contexts, rather than viewing entities in isolation; the prompt can be found in our open-source codebase.


To validate the reliability of our approach, we additionally recruited two researchers specializing in Natural Language Processing (NLP) to conduct a manual secondary evaluation of the LLM-assessed samples. The results reveal that the entity subsets extracted by the LLM strictly subsume those identified by human evaluators. Although the LLM occasionally extracts redundant entities—leading to a generally lower overall score—anonymizing supplementary entities introduces no additional privacy compromises. Furthermore, the downstream utility still maintains a clear advantage over naive baselines that anonymize all recognized entities. Therefore, we conclude that adopting the LLM-based evaluation paradigm is sound.

\section{Proof of Semantic Security}
\label{ap:2}
We employ the Game-Based Proof technique standard in modern cryptography. 

Game $G_0$ (Real World): The challenger selects two entities $e_0,e_1$ such that ${\Pi_{\Gamma}}\left(e_0\right)={\Pi_{\Gamma}}\left(e_1\right)=c$ and $\Psi\left(e\right)\geq\tau$. A random bit $b\ \in\{0,1\}$ is chosen, and the system outputs $T^\prime\ =\ {\Pi_{\Gamma}}_{{e}_b}\left({T}\right)$. The adversary $\mathcal{A}$ guesses $b^{\prime}$.
\begin{equation}
	\Pr\left[S_0\right]=\Pr\left[b^\prime=b\right],
\end{equation}

Game $G_1$ (Deterministic Replacement): The mechanism is modified to directly replace the entity slot with the label c without processing the specific entity features, relying on the deterministic nature of $\Pi$. Since $\Pi$ is deterministic for a given input and context, the output distribution remains bitwise identical to $G_0$.
\begin{equation}
	\Pr\left[S_1\right]=\Pr\left[S_0\right],
\end{equation}

Game $G_2$ (Ideal/Blind World): The challenger generates the output using only the template $T_{template}$ and the class labels $c$, independent of b. The view of $\mathcal{A}$ contains no information about b. Thus, the probability of guessing b is purely random.
\begin{equation}
	\Pr\left[S_2\right]=\frac{1}{2},
\end{equation}

Conclusion: The advantage is $|\Pr[S_0] - 12| = |\Pr[S_2] - 12| = 0$. This proves absolute semantic indistinguishability for generalized entities.
\section{Introduction to Experimental Baseline}
\label{ap:4}
\begin{itemize}
	\item \textbf{Redact} obtains the anonymized text by performing a generalized mapping replacement on the identified and extracted entities based on their labels.
	\item \textbf{Paraphrase} utilizes the capabilities of LLMs to extract relevant and important components from the retrieved context. Less important parts can be filtered out, and certain sentences may require rewriting~\cite{jainbaseline2023}.
	\item \textbf{LPRAG} The LPRAG method employs three modules—a word perturbation module, a phrase perturbation module, and a number perturbation module—to perform local differential privacy perturbation~\cite{he2025mitigating}. During our reproduction, we allocated a global privacy budget of 5.0.
\end{itemize}
\section*{GenAI Usage Disclosure}
Generative AI tools were employed strictly for auxiliary tasks during this study. The fundamental research contributions, algorithmic design, empirical analysis, and manuscript drafting were executed entirely by the authors without AI involvement.

\textbf{Literature Search and Discovery.} During the preliminary research stage, we utilized Gemini Deep Research to navigate the existing literature on privacy preservation and data anonymization within Retrieval-Augmented Generation (RAG) frameworks. This tool facilitated the discovery of the study by \citet{he2025mitigating} regarding locally private entity perturbation. Engaging with their findings provided crucial insights into current state-of-the-art approaches in this domain, thereby establishing a robust comparative baseline and informing the development of our context-aware anonymization methodology. Beyond this initial discovery phase, all literature synthesis and critical analysis were performed manually. 

\textbf{Data Formatting.} Generative AI was used to translate CSV files into LaTeX tables for manuscript integration. The underlying datasets, along with all associated analyses and scientific interpretations, are exclusively human-authored.
\bibliographystyle{ACM-Reference-Format}
\bibliography{sample-base}
\end{document}